\documentclass[twocolumn,english,pre,aps,superscriptaddress]{revtex4-2}
\usepackage[utf8]{inputenc}
\usepackage{amsmath}
\usepackage{amssymb}
\usepackage{graphicx}

\makeatletter
\usepackage{color}
\usepackage{babel}

\usepackage[usenames,dvipsnames]{xcolor}

\makeatother

\usepackage{babel}
\begin{document}
\title{Nonorthogonal-state erasure as the resource behind apparent second-law
violations}
\author{Xinshu Xia}
\address{Graduate School of China Academy of Engineering Physics, No. 10 Xibeiwang
East Road, Haidian District, Beijing, 100193, China}
\author{Hui Hui Qin}
\address{Graduate School of China Academy of Engineering Physics, No. 10 Xibeiwang
East Road, Haidian District, Beijing, 100193, China}
\address{Department of Sciences, Hangzhou Dianzi University, Hangzhou 310018,
China}
\author{Yu-Han Ma}
\address{Graduate School of China Academy of Engineering Physics, No. 10 Xibeiwang
East Road, Haidian District, Beijing, 100193, China}
\author{Chang-Pu Sun}
\email{suncp@gscaep.ac.cn}

\address{Graduate School of China Academy of Engineering Physics, No. 10 Xibeiwang
East Road, Haidian District, Beijing, 100193, China}
\author{Hui Dong}
\email{hdong@gscaep.ac.cn}

\address{Graduate School of China Academy of Engineering Physics, No. 10 Xibeiwang
East Road, Haidian District, Beijing, 100193, China}
\date{\today}
\begin{abstract}
Perfect deterministic distinguishing of nonorthogonal quantum states
is forbidden by the linear and unitary structure of quantum mechanics.
It has often been assumed that, if such distinguishing were available,
it would be the resource enabling work extraction from a single heat
bath. We show that this expectation identifies the wrong thermodynamic
operation and prove such hypothetical operation increases, rather
than decreases, the joint entropy of system and detector. The entropy-decreasing
resource is instead the inverse operation, which we call nonorthogonal-state
erasure. Reanalyzing a Peres-type Szilard engine, we show that the
apparent extracted work $W_{\mathrm{ext}}=0.2766k_{\mathrm{B}}T$
for an equal mixture of an atomic ensemble with spin state $\left|\uparrow\right\rangle $
and $\left|\rightarrow\right\rangle $. Thus the apparent second-law
violation is supplied not by nonorthogonal-state distinguishing, but
by a nonorthogonal quantum state erasure. 
\end{abstract}
\maketitle

\section{Introduction}

Quantum measurement occupies a central position in the relation between
quantum mechanics and thermodynamics, especially in thermodynamic
processes involving information acquisition and feedback control,
e.g., Maxwell-demon-type cycles \citet{LeffRex1990MaxwellsDemon,Peres2002,Sagawa2012ThermodynamicsInformationProcessing}.
Measurement is not merely a passive observation, but a physical process
that creates a record of a microscopic state and enables subsequent
operations conditioned on that record to extract work from a heat
bath \citet{Quan2006,Dong2011,Cai2012}. In the quantum regime, such
a measurement correlates states of the system with distinguishable
states of a detector \citet{Peres2002}. If the system states are
$\{|\psi_{i}\rangle\}$ and the detector is initially in a fixed state
$|D\rangle$, the distinguishing operation $\mathcal{D}$ is expressed
explicitly as 
\begin{equation}
\mathcal{D}:\left|\psi_{i}\right\rangle \otimes\left|D\right\rangle \mapsto\left|\psi_{i}\right\rangle \otimes\left|D_{i}\right\rangle ,
\end{equation}
where the detector states $|D_{i}\rangle$ encode the measurement
outcome. A faithful measurement record requires the detector states
to be distinguishable, namely $\langle D_{i}|D_{j}\rangle=\delta_{ij}$.
For mutually orthogonal system states $\langle\psi_{i}|\psi_{j}\rangle=\delta_{ij}$,
the distinguishing process preserves all inner products and can therefore
be embedded in a unitary evolution on the joint system-detector Hilbert
space. It is reversible at the level of the total system and does
not provide an independent entropy-decreasing resource.

However, the situation changes qualitatively for nonorthogonal states
$\langle\psi_{i}|\psi_{j}\rangle\neq\delta_{ij}$ \citet{Peres2002,Zurek_2007,Scarani2005}.
The thermodynamic question is then not only whether such an operation
is allowed, but what resource it would represent if it were supplied
hypothetically. A prevailing expectation is that, if a device could
perfectly distinguish nonorthogonal states, then it could be used
to extract work from a single heat bath \citet{Peres2002,Maruyama2009,polo2024thermodynamic}.
In this Letter, we show that this expectation misidentifies the entropy-decreasing
step. We prove here the distinguishing operation itself always increases
the joint entropy. Instead, the relevant resource is its inverse operation,
non-orthogonal-state erasure (NOSE) $\mathcal{E}:\left|\psi_{i}\right\rangle \otimes\left|D_{i}\right\rangle \mapsto\left|\psi_{i}\right\rangle \otimes\left|D\right\rangle $,
which reduces entropy and can therefore mimic a second-law violation
if its thermodynamic cost is not paid.

\section{NONORTHOGONAL-STATE ERASURE AS AN ENTROPY-DECREASING RESOURCE}

We consider an ensemble of normalized, yet not necessarily orthogonal
states $\rho_{S}=\sum^{m}_{i=1}p_{i}\left|\psi_{i}\right\rangle \left\langle \psi_{i}\right|,$with
$\sum p_{i}=1$. Before the hypothetical distinguishing operation,
the joint system-detector state is 
\begin{equation}
\rho_{\mathrm{bef}}=\rho_{S}\otimes\left|D\right\rangle \left\langle D\right|.\label{eq:rho_bef}
\end{equation}
After the operation, it becomes 
\begin{equation}
\rho_{\mathrm{aft}}=\sum^{m}_{i=1}p_{i}\left|\psi_{i}\right\rangle \left\langle \psi_{i}\right|\otimes\left|D_{i}\right\rangle \left\langle D_{i}\right|.\label{eq:rho_pre}
\end{equation}
Because the detector labels are orthogonal, the nonzero eigenvalues
of $\rho_{\mathrm{aft}}$ are simply $\{p_{i}\}^{m}_{i=1}$, resulting
in the entropy $S_{\mathrm{aft}}=-\sum^{m}_{i=1}p_{i}\ln p_{i}$ for
the total system after the operation. On the other hand, the entropy
before the operation is $S_{\mathrm{bef}}=-\mathrm{Tr}[\rho_{\mathrm{bef}}\ln\rho_{\mathrm{bef}}]=-\mathrm{Tr}[\rho_{S}\ln\rho_{S}]$.
The entropy satisfies the inequality 
\begin{equation}
S_{\mathrm{bef}}\leq S_{\mathrm{aft}},\label{eq:nogotherem}
\end{equation}
with equality only when the states carrying nonzero probabilities
are mutually orthogonal. A proof of Eq. (\ref{eq:nogotherem}) based
on the Schur-Horn theorem \citet{Horn1954,Devadas2015} is given in
the Appendix A. The entropy change $\Delta S_{\mathcal{D}}=S_{\mathrm{aft}}-S_{\mathrm{bef}}$
is non-negative, i.e., $\Delta S_{\mathcal{D}}\geq0$.

\begin{figure}
\includegraphics{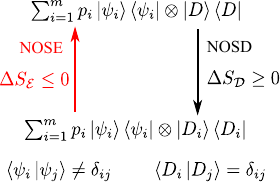} \caption{The relation between entropy change and non-orthogonal state distinguishing
(erasure). NOSD(E): Non-orthogonal-state distinguishing (erasure).}
\label{fig:era_dis} 
\end{figure}

Equation (\ref{eq:nogotherem}) is the central result for entropy
change caused by the hypothetical non-orthogonal-state distinguishing.
The nonorthogonal state distinguishing operation creates an orthogonal
classical label and converts indistinguishable quantum overlap into
distinguishable classical labels, and therefore increases the joint
entropy. Thus the distinguishing operation itself is not an entropy-decreasing
resource. Its inverse operation has the opposite character: NOSE maps
the state in Eq. (\ref{eq:rho_pre}) back to the state in Eq. (\ref{eq:rho_bef}).
Its entropy change $\Delta S_{\mathcal{E}}=-\Delta S_{\mathcal{D}}$
is non-positive. This operation erases the orthogonal register labels
$\left|D_{i}\right\rangle $ while preserving the corresponding nonorthogonal
states $\left|\psi_{i}\right\rangle $. It changes inner products,
two initially orthogonal joint states $\left|\psi_{i}\right\rangle \otimes\left|D_{i}\right\rangle $,$\left|\psi_{j}\right\rangle \otimes\left|D_{j}\right\rangle $
are mapped to states $\left|\psi_{i}\right\rangle \otimes\left|D\right\rangle $,$\left|\psi_{j}\right\rangle \otimes\left|D\right\rangle $
with overlap $\left\langle \psi_{i}\right.\left|\psi_{j}\right\rangle $.
For nonorthogonal $\left|\psi_{i}\right\rangle $ and $\left|\psi_{j}\right\rangle $,
this operation $\mathcal{E}$ cannot be achieved by a unitary or isometric
evolution and therefore supplies negative entropy. We summarize the
relation between the entropy change and non-orthogonal state distinguishing
(erasure) with the diagram in Fig. \ref{fig:era_dis}. Non-orthogonal
state distinguishing describes the evolution from the state $\sum^{m}_{i=1}p_{i}\left|\psi_{i}\right\rangle \left\langle \psi_{i}\right|\otimes\left|D\right\rangle \left\langle D\right|$
to the state $\sum^{m}_{i=1}p_{i}\left|\psi_{i}\right\rangle \left\langle \psi_{i}\right|\otimes\left|D_{i}\right\rangle \left\langle D_{i}\right|$,
which results in the entropy increase. The reverse operation, nonorthogonal-state
erasure, decreases the entropy. If supplied without the compensating
thermodynamic cost quantified in Eq. (\ref{eq:workExtract}), this
operation would provide the resource needed for an apparent second-law
violation.

To make the result explicit, we consider the two-state ensemble, e.g.,
a spin-1/2 system with the orthogonal basis $\left|\uparrow\right\rangle $
and $\left|\downarrow\right\rangle $. The system is initially in
the state $\rho_{S}=[\left|\uparrow\right\rangle \left\langle \uparrow\right|+\left|\rightarrow\right\rangle \left\langle \rightarrow\right|]/2$,
where $\left|\rightarrow\right\rangle =[\left|\uparrow\right\rangle +\left|\downarrow\right\rangle ]/\sqrt{2}$
is not orthogonal with the state $\left|\uparrow\right\rangle $.
The measurement device has the orthogonal state $\left|e\right\rangle $
and $\left|g\right\rangle $ with $\left\langle e\right.\left|g\right\rangle =0$.
The pre-measured state is $\rho_{\mathrm{aft}}=[\left|\uparrow\right\rangle \left\langle \uparrow\right|\otimes\left|e\right\rangle \left\langle e\right|+\left|\rightarrow\right\rangle \left\langle \rightarrow\right|\otimes\left|g\right\rangle \left\langle g\right|]/2$
with the entropy $S_{\mathrm{aft}}=\ln2$, which is larger than that
before measurement, i.e., $S_{\mathrm{bef}}=-\mathrm{Tr}[\rho_{S}\ln\rho_{S}]=-\cos^{2}\theta\ln\cos^{2}\theta-\sin^{2}\theta\ln\sin^{2}\theta$
with $\theta=\pi/8$. The two eigenstates for the density matrix $\rho_{S}$
are $\left|\nearrow\right\rangle =\cos\theta\left|\uparrow\right\rangle +\sin\theta\left|\downarrow\right\rangle $
and $\left|\searrow\right\rangle =-\sin\theta\left|\uparrow\right\rangle +\cos\theta\left|\downarrow\right\rangle $.
The entropy increase associated with distinguishing the two nonorthogonal
states is therefore $\Delta S_{\mathcal{D}}=S_{\mathrm{aft}}-S_{\mathrm{bef}}\backsimeq0.2766$.
The inverse erasure operation decreases entropy by exactly the same
amount, i.e., $\Delta S_{\mathcal{E}}\backsimeq-0.2766$. In thermodynamic
units, the entropy reduction corresponds to a minimum erasure cost
$k_{\mathrm{B}}T\Delta S_{\mathcal{D}}\backsimeq0.2766k_{\mathrm{B}}T$,
which is the same scale as the apparent work gain in the cycle below.

Perfect deterministic distinguishing of nonorthogonal quantum states
is entropy increasing and therefore cannot by itself fuel a single-bath
engine. The entropy-reducing operation is instead its inverse, erasure
of an orthogonal record while preserving the associated nonorthogonal
states. If implemented physically in contact with a heat bath at temperature
$T$, its entropy decrease would have to be compensated by an external
cost at least 
\begin{equation}
W_{\mathrm{cost}}\geq-k_{\mathrm{B}}T\Delta S_{\mathcal{E}}=k_{\mathrm{B}}T\Delta S_{\mathcal{D}}.\label{eq:workExtract}
\end{equation}
If this cost is not paid, the operation itself already contains the
resource needed to violate the second law.

At first sight, this conclusion appears to conflict with Peres-type
proposals \citep{Peres2002,Maruyama2009}, in which nonorthogonal
states seem to assist work extraction from a single heat bath. We
now revisit this representative model and show that the apparent violation
is not caused by nonorthogonal-state distinguishing, but by its inverse
operation, nonorthogonal-state erasure.

\section{Peres' model}

\begin{figure*}
\includegraphics[width=16cm]{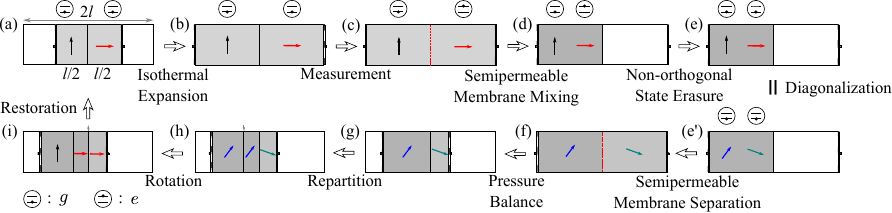}

\caption{Modified Peres' model of the second law violation with non-orthogonal
erasure. (a) The initial state of the setup. A single spin-1/2 atom
with spin states $\left|\uparrow\right\rangle $ and $\left|\rightarrow\right\rangle $
is trapped in a one-dimension chamber with equal probability in the
left and right compartments, illustrated as gray regions. A single
register with initial state $\left|g\right\rangle $ is included as
the memory for the measurement to avoid any Maxwell-demon type second
law violation. (a$\rightarrow$b) Isothermal expansion. The pistons
on the two sides are moved to double the widths. (b$\rightarrow$c)
Measurement. The register state is flipped to its excited state $\left|e\right\rangle $
if the single atom is in the right compartment, and remains on the
ground state if the atom is in the left compartment. (c$\rightarrow$d)
Semipermeable membrane mixing. A semipermeable membrane is inserted
to allow only an atom with spin state $\left|\rightarrow\right\rangle $
to the left compartment only when the register state is $\left|e\right\rangle $.
(d$\rightarrow$e) Non-orthogonal erasure. The state of the register
is restored to the initial state $\left|g\right\rangle $. This operation
is not allowed by unitary evolution. (e$\rightarrow$e') Diagonalization.
The spin state is rewritten with the orthogonal states $\left|\nearrow\right\rangle $
and $\left|\searrow\right\rangle $ represented by the green and blue
arrow. (e'$\rightarrow$f) Semipermeable membrane separation. A new
semipermeable membrane is used to allow only spin state $\left|\searrow\right\rangle $
to the right compartment. (f$\rightarrow$g) Pressure balance. The
two pistons are moved to ensure the total width is $l$ and the pressure
is the same as that in figure (a). (g$\rightarrow$h) Repartition.
An additional wall is inserted to ensure the width of the left compartment
is $l/2$. (h$\rightarrow$i) Rotation. The spin state for the atom
in three regions is rotated accordingly. (i$\rightarrow$a) Restoration.
The wall between the middle and the right compartment is removed and
the two pistons are shifted to the original positions.}
\label{fig:Peres_model} 
\end{figure*}

We illustrate Peres' model in Fig. \ref{fig:Peres_model}. For simplicity,
we consider a single spin-1/2 atom following the standard simplification
used in Szilard-engine models \citep{Szilard1929,Kim2011,Dong2011,Cai2012}.
Initially, the atom stays within half of the one-dimensional chamber
with total width $2l$, and the gas is separated in the middle with
an impenetrable wall, as shown in Fig. \ref{fig:Peres_model}(a).
The spin state of the atom on the left is spin up $\left|\uparrow\right\rangle $,
and the spin state of the atom on the right is $\left|\rightarrow\right\rangle $.
The state of the single atom is $\rho_{a}=\frac{1}{2}[\left|\uparrow\right\rangle \left\langle \uparrow\right|\otimes\rho_{L}(l/2)+\left|\rightarrow\right\rangle \left\langle \rightarrow\right|\otimes\rho_{R}(l/2)]$,
where $\rho_{\alpha}(l)$ is the density matrix for the spatial degree
of freedom with $\alpha=L$ and $R$, and width $l$ as detailed in
the Appendix B. To avoid any possible Maxwell-demon-type paradox,
we add a register bit with the ground state $\left|g\right\rangle $
and excited state $\left|e\right\rangle $. The register acts as a
measuring device and, together with the gas, constitutes the prepared
state $\rho_{\mathrm{tot},a}=\rho_{a}\otimes\left|g\right\rangle \left\langle g\right|$.
The register can be another spin or other degree of freedom of the
spin-1/2 atom. We describe the processes as follows.

\textbf{Isothermal expansion} (a$\rightarrow$b): The two pistons
on the two sides expand isothermally to double the width, with the
end state illustrated in Fig. \ref{fig:Peres_model}(b). The work
extracted in the process is $W_{a\rightarrow b}=k_{\mathrm{B}}T\ln2$
in the large-scale limit $l\gg\lambda_{T}$, where $\lambda_{T}=\hbar\sqrt{2\pi/mk_{B}T}$
is the thermal wavelength and $m$ is the mass of the atom, as detailed
in the Appendix B. The state of total system is $\rho_{\mathrm{tot},b}=\frac{1}{2}[\left|\uparrow\right\rangle \left\langle \uparrow\right|\otimes\rho_{L}(l)+\left|\rightarrow\right\rangle \left\langle \rightarrow\right|\otimes\rho_{R}(l)]\otimes\left|g\right\rangle \left\langle g\right|$.

\textbf{Measurement }(b$\rightarrow$c): The measurement is performed
with a unitary transformation $U_{bc}=I\otimes[P_{L}(l)\otimes\left|g\right\rangle \left\langle g\right|+P_{R}(l)\otimes(\left|e\right\rangle \left\langle g\right|+\left|g\right\rangle \left\langle e\right|)]$,
resulting in the final state $\rho_{\mathrm{tot},c}=\frac{1}{2}[\left|\uparrow\right\rangle \left\langle \uparrow\right|\otimes\rho_{L}(l)\otimes\left|g\right\rangle \left\langle g\right|+\left|\rightarrow\right\rangle \left\langle \rightarrow\right|\otimes\rho_{R}(l)\otimes\left|e\right\rangle \left\langle e\right|]$,
where $P_{L}(l)$ and $P_{R}(l)$ are the projection operators for
each spatial region, as detailed in the Appendix B. Crucially, since
the spatial wavefunctions in left and right chambers are orthogonal,
the measurement is effectively performed in an orthogonal basis of
the joint spin-spatial Hilbert space.

This observation resolves the apparent tension with our no-go theorem.
In the original conception, the spin states $|\uparrow\rangle$ and
$|\rightarrow\rangle$ are non-orthogonal, which led to the belief
that distinguishing them without disturbing the spatial degree of
freedom constitutes a genuine non-orthogonal measurement. However,
the measurement process explicitly couples spin to the orthogonal
spatial projectors $P_{L}(l)$ and $P_{R}(l)$. Consequently, the
measurement is not an implementation of the non-unitary evolution
$\mathcal{D}$ assumed in Eq. (\ref{eq:rho_pre}), but rather a standard
projective measurement on the composite system. The entropy increase
associated with a hypothetical non-orthogonal distinction is therefore
never triggered in this step. No work is done in this process.

\textbf{Semipermeable membrane mixing }(c$\rightarrow$d): A semipermeable
membrane replaces the wall in the middle. The membrane is designed
to detect the state of the register and applies the operation to allow
the gas to pass only through from the right to the left. The state
is $\rho_{\mathrm{tot},d}=\frac{1}{2}[\left|\uparrow\right\rangle \left\langle \uparrow\right|\otimes\rho_{L}(l)\otimes\left|g\right\rangle \left\langle g\right|+\left|\rightarrow\right\rangle \left\langle \rightarrow\right|\otimes\rho_{L}(l)\otimes\left|e\right\rangle \left\langle e\right|]$.
No work is done.

\textbf{Non-orthogonal state erasure }(d$\rightarrow$e): The information
in the register is erased. The state is $\rho_{\mathrm{tot},e}=\frac{1}{2}[\left|\uparrow\right\rangle \left\langle \uparrow\right|+\left|\rightarrow\right\rangle \left\langle \rightarrow\right|]\otimes\rho_{L}(l)\otimes\left|g\right\rangle \left\langle g\right|$.
No work is done by the mechanical degrees of freedom, but the entropy
decreases. Ignoring the spatial degree of freedom, the entropy before
the erasure is $S_{d}=\ln2$, while the entropy after is $S_{e}=-\mathrm{Tr}[\rho_{\mathrm{tot},e}\ln\rho_{\mathrm{tot},e}]=0.416$.

\textbf{Diagonalization }(e$\rightarrow$e'): The state is rewritten
as $\rho_{\mathrm{tot},e'}=[p_{\nearrow}\left|\nearrow\right\rangle \left\langle \nearrow\right|+p_{\searrow}\left|\searrow\right\rangle \left\langle \searrow\right|]\otimes\rho_{L}(l)\otimes\left|g\right\rangle \left\langle g\right|$
with $p_{\nearrow}+p_{\searrow}=1$, where the exact forms of $\left|\nearrow\right\rangle $
and $\left|\searrow\right\rangle $ are given in the example above,
and $p_{\nearrow}=0.854$ is the probability that the atom is in the
state $\left|\nearrow\right\rangle $. No work is done.

\textbf{Semipermeable membrane separation }(e'$\rightarrow$f): Another
semipermeable membrane is used to separate the gas into the two sides
with the state in Fig. \ref{fig:Peres_model}(f) as $\rho_{\mathrm{tot},f}=[p_{\nearrow}\left|\nearrow\right\rangle \left\langle \nearrow\right|\otimes\rho_{L}(l)+p_{\searrow}\left|\searrow\right\rangle \left\langle \searrow\right|\otimes\rho_{R}(l)]\otimes\left|g\right\rangle \left\langle g\right|$.
The separation is done by a conditional operation similar to that
in the process $c\rightarrow d$. No work is done.

\textbf{Pressure balance }(f$\rightarrow$g): In the large-scale limit
$l\gg\lambda_{T}$, the pressure is inversely proportional to the
width, as detailed in the Appendix B. Hence, the two pistons are moved
to make the widths of the gas as $p_{\nearrow}l$ on the left and
$p_{\searrow}l$ on the right as illustrated in Fig. \ref{fig:Peres_model}(g).
The state is $\rho_{\mathrm{tot},g}=[p_{\nearrow}\left|\nearrow\right\rangle \left\langle \nearrow\right|\otimes\rho_{L}(p_{\nearrow}l)+p_{\searrow}\left|\searrow\right\rangle \left\langle \searrow\right|\otimes\rho_{R}(p_{\searrow}l)]\otimes\left|g\right\rangle \left\langle g\right|$.
The amount of work done is $W_{f\rightarrow g}=k_{B}T(p_{\nearrow}\ln p_{\nearrow}+p_{\searrow}\ln p_{\searrow})=-0.416k_{B}T$.

\textbf{Repartition }(g$\rightarrow$h): A new wall is inserted in
the left compartment so that the left gas width is $l/2$ and the
gas width of the middle part as $(p_{\nearrow}-1/2)l$. The state
in Fig. \ref{fig:Peres_model}(h) is $\rho_{\mathrm{tot},h}=[\frac{1}{2}\left|\nearrow\right\rangle \left\langle \nearrow\right|\otimes\rho_{L}(l/2)+(p_{\nearrow}-\frac{1}{2})\left|\nearrow\right\rangle \left\langle \nearrow\right|\otimes\rho_{M}((p_{\nearrow}-\frac{1}{2})l)+p_{\searrow}\left|\searrow\right\rangle \left\langle \searrow\right|\otimes\rho_{R}((1-p_{\nearrow})l)]\otimes\left|g\right\rangle \left\langle g\right|$.
No work is done.

\textbf{Rotation }(h$\rightarrow$i): Three different pulses are applied
in three regions to rotate the spin state to $\left|\uparrow\right\rangle $
for the left compartment, and to $\left|\rightarrow\right\rangle $
in the middle and right compartment, with a unitary transformation
$U_{hi}=[(\left|\uparrow\right\rangle \left\langle \nearrow\right|+\left|\downarrow\right\rangle \left\langle \searrow\right|)\otimes P_{L}(\frac{l}{2})+(\left|\rightarrow\right\rangle \left\langle \nearrow\right|+\left|\leftarrow\right\rangle \left\langle \searrow\right|)\otimes P_{M}((p_{\nearrow}-\frac{1}{2})l)+(\left|\rightarrow\right\rangle \left\langle \searrow\right|+\left|\leftarrow\right\rangle \left\langle \nearrow\right|)\otimes P_{R}(p_{\searrow}l)]\otimes I$
resulting in the state of the system as $\rho_{\mathrm{tot},i}=[\frac{1}{2}\left|\uparrow\right\rangle \left\langle \uparrow\right|\otimes\rho_{L}(l/2)+(p_{\nearrow}-\frac{1}{2})\left|\rightarrow\right\rangle \left\langle \rightarrow\right|\otimes\rho_{M}((p_{\nearrow}-\frac{1}{2})l)+p_{\searrow}\left|\rightarrow\right\rangle \left\langle \rightarrow\right|\otimes\rho_{R}(p_{\searrow}l)]\otimes\left|g\right\rangle \left\langle g\right|$,
where $\left|\leftarrow\right\rangle =[\left|\uparrow\right\rangle -\left|\downarrow\right\rangle ]/\sqrt{2}$
is not orthogonal with the state $\left|\uparrow\right\rangle $.
Here $P_{L}(\frac{l}{2})$, $P_{M}((p_{\nearrow}-\frac{1}{2})l)$
and $P_{R}(p_{\searrow}l)$ are the projection operators for each
spatial region, as detailed in the Appendix B. No work is done.

\textbf{Restoration }(i$\rightarrow$a): The wall separating the middle
and the right compartment is removed to form two compartments with
equal width $l/2$ and the state is $\rho_{\mathrm{tot},i}=[\frac{1}{2}\left|\uparrow\right\rangle \left\langle \uparrow\right|\otimes\rho_{L}(l/2)+\frac{1}{2}\left|\rightarrow\right\rangle \left\langle \rightarrow\right|\otimes\rho_{R}(l/2)]\otimes\left|g\right\rangle \left\langle g\right|$.
The system is restored to the initial state illustrated in Fig. \ref{fig:Peres_model}(a).
In this process, no work is done.

The cycle differs from Peres' original construction in how the nonorthogonal-state
operation is identified. The measurement step is not nonorthogonal-state
distinguishing once spatial degrees of freedom are included; the extra
entropy-reducing step is nonorthogonal-state erasure.

To quantify the apparent violation, we evaluate the total work extracted
as 
\begin{equation}
W_{\mathrm{ext}}=W_{a\rightarrow b}+W_{f\rightarrow g}=k_{\mathrm{B}}T[\ln2-S(\rho_{S})]=0.2766k_{\mathrm{B}}T.
\end{equation}
Here, the engine appears to extract work from a single bath with temperature
$T$, which would violate the second law only if the erasure cost
were not paid. Comparing this expression with Eq. (\ref{eq:workExtract}),
the apparent gain is exactly canceled by the minimum cost of the entropy-decreasing
nonorthogonal-state erasure operation. The second-law violation is
therefore not caused by free nonorthogonal-state distinguishability;
it is caused by treating the forbidden erasure resource as costless.

Indeed, two basic structural principles of quantum mechanics, the
linearity and unitarity, have prohibited such perfect deterministic
distinguishing and erasure of nonorthogonal quantum states \citet{Zurek_2007,Scarani2005}.
This can be seen from a simple example. In a measurement, the device
correlates two system states $\left|\alpha\right\rangle $ and $\left|\beta\right\rangle $
with two detector states $\left|D_{\alpha}\right\rangle $ and $\left|D_{\beta}\right\rangle $,
unitarity requires $\left\langle \alpha\right.\left|\beta\right\rangle =\left\langle \alpha\right.\left|\beta\right\rangle \left\langle D_{\alpha}\right.\left|D_{\beta}\right\rangle $.
Hence, whenever $\left\langle \alpha\right.\left|\beta\right\rangle \neq0$,
the detector states cannot be orthogonal. This obstruction is the
same inner-product preservation behind the no-cloning theorem \citet{Wootters1982,Scarani2005},
where the norm preservation of unitary evolution is used \citet{Peres2002}.

\section{Conclusion}

In summary, we have shown that the thermodynamic resource in Peres-type
nonorthogonal-state engines has been misidentified. A hypothetical
operation distinguishing nonorthogonal quantum states is entropy increasing
for the joint system and measurement device, and therefore cannot
by itself fuel work extraction from a single heat bath. The entropy-reducing
operation is instead its inverse, namely the erasure of an orthogonal
record while preserving the corresponding nonorthogonal states. Reexamining
a Peres-type Szilard engine, we find that the apparent extracted work
is exactly matched by the minimum cost of this nonorthogonal-state
erasure. This identifies NOSE, not nonorthogonal-state distinguishability,
as the hidden resource behind the apparent second-law violation.
\begin{acknowledgments}
HD thanks D.Z. Xu for helpful discussion on the early stage of the
current work. This work is supported by the National Natural Science
Foundation of China (Grant No. 12088101, No. U2230203, and No. U2330401),
and the Quantum Science and Technology-National Science and Technology
Major Project (Grant No. 2023ZD0300700). 
\end{acknowledgments}

\section*{Appendix A: Proof of the Entropy Inequality in the No-Go Statement}

Consider a quantum system described by a $d$-dimensional Hilbert
space $V_{S}$. The system is initially prepared in a mixed state
$\rho_{S}=\sum^{m}_{i=1}p_{i}\left|\psi_{i}\right\rangle \left\langle \psi_{i}\right|$,
where each $\left|\psi_{i}\right\rangle $ is normalized ($\left\langle \psi_{i}\right.\left|\psi_{i}\right\rangle =1$),
and $\sum^{m}_{i=1}p_{i}=1$. The measuring apparatus is initially
in a pure state $\left|D\right\rangle $. Hence the total initial
state of system plus apparatus is 
\begin{equation}
\rho_{\mathrm{bef}}=\rho_{S}\otimes\left|D\right\rangle \left\langle D\right|.
\end{equation}

The apparatus has $m$ distinguishable orthogonal states $\left\{ \left|D_{i}\right\rangle \right\} ^{m}_{i=1}$.
After the measurement, the total state becomes $\rho_{\mathrm{aft}}=\sum^{m}_{i=1}p_{i}\left|\psi_{i}\right\rangle \left\langle \psi_{i}\right|\otimes\left|D_{i}\right\rangle \left\langle D_{i}\right|$.
The von Neumann entropies of these two states are 
\begin{equation}
S_{\mathrm{bef}}=-\mathrm{Tr}[\rho_{S}\ln\rho_{S}],S_{\mathrm{aft}}=-\sum^{m}_{i=1}p_{i}\ln p_{i}.
\end{equation}
We shall prove that $S_{\mathrm{bef}}\leq S_{\mathrm{aft}}$, with
equality if and only if $\left\langle \psi_{i}\right.\left|\psi_{j}\right\rangle =\delta_{ij}$.

The density operator $\rho_{S}$ is Hermitian and can be spectrally
decomposed as 
\begin{equation}
\rho_{S}=\sum^{d}_{i=1}\lambda_{i}\left|\varphi_{i}\right\rangle \left\langle \varphi_{i}\right|,
\end{equation}
where $\lambda_{i}\geq0,\sum^{d}_{i=1}\lambda_{i}=1$, and $\left\{ \left|\varphi_{i}\right\rangle \right\} ^{d}_{i=1}$
is an orthonormal basis of $V_{S}$.

Now introduce an $m$-dimensional Hilbert space $V_{m}$ with orthonormal
basis $\left\{ \left|i\right\rangle \right\} ^{m}_{i=1}$. Define
a linear operator $A:V_{m}\rightarrow V_{S},A=\sum^{m}_{i=1}\sqrt{p_{i}}\left|\psi_{i}\right\rangle \left\langle i\right|$.
Its adjoint acts as $A^{\dagger}:V_{S}\rightarrow V_{m},A^{\dagger}=\sum^{m}_{i=1}\sqrt{p_{i}}\left|i\right\rangle \left\langle \psi_{i}\right|$.
Then we have $AA^{\dagger}:V_{S}\rightarrow V_{S},AA^{\dagger}=\sum^{m}_{i=1}A\left|i\right\rangle \left\langle i\right|A^{\dagger}=\rho_{S}$.
Define the auxiliary operator $\sigma=A^{\dagger}A:V_{m}\rightarrow V_{m},$
\begin{equation}
\sigma=\sum^{m}_{i,j=1}\sqrt{p_{i}p_{j}}\left\langle \psi_{i}\right.\left|\psi_{j}\right\rangle \left|i\right\rangle \left\langle j\right|,
\end{equation}
which is positive semidefinite and Hermitian. Its diagonal elements
are $\left\langle i\right|\sigma\left|i\right\rangle =p_{i}$, and
$\mathrm{Tr}[\sigma]=1$.

The operators $AA^{\dagger}$ and $A^{\dagger}A$ have the same non-zero
eigenvalues with the same multiplicities. Since $\rho_{S}=AA^{\dagger}$
has eigenvalues $\{\lambda_{i}\}^{d}_{i=1}$, the auxiliary operator
$\sigma=A^{\dagger}A$ has exactly the same eigenvalues $\{\lambda_{i}\}^{d}_{i=1}$,
the rest being zero. Using $0\ln0=0$, the von Neumann entropy is
determined solely by the non-zero eigenvalues, hence 
\begin{equation}
S(\sigma)=-\sum^{d}_{i=1}\lambda_{i}\ln\lambda_{i}=S(\rho_{S}).
\end{equation}

The Hermitian properties of the auxiliary operator $\sigma$ assure
that the eigenvalues of the operator $\sigma$, denoted as $\overrightarrow{\mu}$,
are real numbers, namely $\overrightarrow{\mu}\in\mathbb{R}^{m}$.
One can arrange these eigenvalues in the descending order $\mu_{1}\geq\mu_{2}\geq\cdots\geq\mu_{m}$.
The Schur-Horn theorem \citep{Horn1954,Devadas2015} limits the relation
between the eigenvalues $\overrightarrow{\mu}$ and the diagonal elements
$\overrightarrow{p}$ on the basis $\left\{ \left|i\right\rangle \right\} ^{m}_{i=1}$
as follows,

\begin{equation}
\begin{cases}
\sum^{k}_{i=1}p_{i}\leq\sum^{k}_{i=1}\mu_{i}, & 1\leq k\leq m-1\\
\sum^{m}_{i=1}p_{i}=\sum^{m}_{i=1}\mu_{i}
\end{cases}.
\end{equation}
which is typically known as the majorization $\overrightarrow{\mu}\succ\overrightarrow{p}$.

With the eigenvalues $\overrightarrow{\mu}$ of the auxiliary operator
$\sigma$, we obtain the entropy before measurement as $S_{\mathrm{bef}}=-\sum^{m}_{i=1}\mu_{i}\ln\mu_{i}$.
The Schur-concave property of the function $f\left(\overrightarrow{x}\right)=-\sum_{i}x_{i}\ln x_{i}$
immediately results in a relation 
\begin{equation}
S_{\mathrm{bef}}\leq S_{\mathrm{aft}}.
\end{equation}

\section*{Appendix B: The spatial part of the Peres' model}

A single particle confined to a spatial region with the width $l$
can be described by the one-dimensional infinite potential well model.
The potential $\mathcal{V}(x)$ is 
\begin{equation}
\mathcal{V}(x)=\left\{ \begin{array}{cc}
0, & 0<x<l\\
+\infty, & \mathrm{otherwise}
\end{array}\right..
\end{equation}
with stationary states $\left|\psi_{n}\right\rangle $, whose eigenvalues
$E_{n}$ and eigenfunctions $\psi_{n}(x)$ given by 
\[
E_{n}=\frac{n^{2}\pi^{2}\hbar^{2}}{2ml^{2}},
\]
\begin{equation}
\psi_{n}(x)=\left\langle x\right.\left|\psi_{n}\right\rangle =\left\{ \begin{array}{cc}
\sqrt{\frac{2}{l}}\sin(\frac{n\pi}{l}x), & 0<x<l\\
0, & \mathrm{otherwise}
\end{array}\right..
\end{equation}

At thermal equilibrium with a heat bath at temperature $T$, the density
matrix is 
\begin{equation}
\rho_{\mathrm{eq}}(l)=\frac{\exp(-\beta H)}{Z}=\frac{1}{Z}\sum^{+\infty}_{n=1}\exp(-\beta E_{n})\left|\psi_{n}\right\rangle \left\langle \psi_{n}\right|,
\end{equation}
where $\beta=1/k_{B}T$ is the inverse temperature, and $Z=\mathrm{Tr}[\exp(-\beta H)]$
is the partition function given as 
\begin{equation}
Z=\sum^{+\infty}_{n=1}\exp(-\frac{\pi\lambda^{2}_{T}}{4l^{2}}n^{2}),
\end{equation}
where $\lambda_{T}=\hbar\sqrt{(2\pi\beta)/m}$ is the thermal wavelength.

According to the free energy $F=-(\ln Z)/\beta$, the pressure $f=-(\partial F/\partial l)_{T}$
exerted on the wall is given by $f=2\overline{E}/l$, where $\overline{E}=-\partial\ln Z/\partial\beta$
is the internal energy. Under an isothermal process, the right wall
is slowly pulled from $x=l_{1}$ to $x=l_{2}$, with the output work
$W^{\mathrm{ex}}$ given as 
\begin{equation}
W^{\mathrm{ex}}=\int^{l_{2}}_{l_{1}}f\mathrm{d}l=F(l_{2})-F(l_{1})=\frac{1}{\beta}\ln\frac{Z(l_{2})}{Z(l_{1})}.
\end{equation}
In the large-scale limit $l\gg\lambda_{T}$, the energy level spacing
is small, and the sum approximates an integral 
\begin{equation}
Z|_{l\gg\lambda_{T}}=\int^{+\infty}_{0}\exp(-\frac{\pi\lambda^{2}_{T}}{4l^{2}}n^{2})\mathrm{d}n=\frac{l}{\lambda_{T}},
\end{equation}
which simplifies the output work as 
\begin{equation}
W^{\mathrm{ex}}|_{l\gg\lambda_{T}}=\frac{1}{\beta}\ln\frac{l_{2}}{l_{1}}.
\end{equation}

Inserting a partition into a region like repartition process (\textbf{$g\rightarrow h$})
is equivalent in the model to adding a delta potential in the one-dimensional
infinite potential well. Consider a one-dimensional infinite potential
well with the width $l=l_{L}+l_{R}$ ranging from $-l_{L}<x<l_{R}$
with a delta potential inserted at $x=0$, 
\begin{equation}
\mathcal{V}(x)=\left\{ \begin{array}{cc}
0, & -l_{L}<x<0,0<x<l_{R}\\
+\infty, & \mathrm{otherwise}
\end{array}\right..
\end{equation}
The stationary states split into left and right sets $\left|\psi^{L}_{n}\right\rangle $
and $\left|\psi^{R}_{n}\right\rangle $, with corresponding eigenvalues
$E^{L}_{n},E^{R}_{n}$ and eigenfunctions $\psi^{L}_{n}(x),\psi^{R}_{n}(x)$
given by 
\[
E^{L}_{n}=\frac{n^{2}\pi^{2}\hbar^{2}}{2ml^{2}_{L}},E^{R}_{n}=\frac{n^{2}\pi^{2}\hbar^{2}}{2ml^{2}_{R}},
\]
\[
\psi^{L}_{n}(x)=\left\langle x\right.\left|\psi^{L}_{n}\right\rangle =\left\{ \begin{array}{cc}
\sqrt{\frac{2}{l_{L}}}\sin(\frac{n\pi}{l_{L}}x), & -l_{L}<x<0\\
0, & \mathrm{otherwise}
\end{array}\right.,
\]
\begin{equation}
\psi^{R}_{n}(x)=\left\langle x\right.\left|\psi^{R}_{n}\right\rangle =\left\{ \begin{array}{cc}
\sqrt{\frac{2}{l_{R}}}\sin(\frac{n\pi}{l_{R}}x), & 0<x<l_{R}\\
0, & \mathrm{otherwise}
\end{array}\right..
\end{equation}

If the system is in thermal equilibrium with a heat source at temperature
$T$, the density matrix of the system is 
\begin{align}
\rho_{\mathrm{eq}} & (l)=\frac{\exp(-\beta H)}{Z}=\frac{1}{Z}[\sum^{+\infty}_{n=1}\exp(-\beta E^{L}_{n})\left|\psi^{L}_{n}\right\rangle \left\langle \psi^{L}_{n}\right|\nonumber \\
 & +\sum^{+\infty}_{n=1}\exp(-\beta E^{R}_{n})\left|\psi^{R}_{n}\right\rangle \left\langle \psi^{R}_{n}\right|],
\end{align}
where $Z=\mathrm{Tr}[\exp(-\beta H)]$ is the partition function.
Defining regional equilibrium state density matrices $\rho^{L}_{\mathrm{eq}}(l_{L}),\rho^{R}_{\mathrm{eq}}(l_{R})$
as 
\[
\rho^{L}_{\mathrm{eq}}(l_{L})=\frac{1}{Z_{L}}\sum^{+\infty}_{n=1}\exp(-\beta E^{L}_{n})\left|\psi^{L}_{n}\right\rangle \left\langle \psi^{L}_{n}\right|,
\]
\begin{equation}
\rho^{R}_{\mathrm{eq}}(l_{R})=\frac{1}{Z_{R}}\sum^{+\infty}_{n=1}\exp(-\beta E^{R}_{n})\left|\psi^{R}_{n}\right\rangle \left\langle \psi^{R}_{n}\right|,
\end{equation}
where the regional partition functions $Z_{L}$ and $Z_{R}$ are 
\[
Z_{L}=\sum^{+\infty}_{n=1}\exp(-\beta E^{L}_{n})=\sum^{+\infty}_{n=1}\exp(-\frac{\pi\lambda^{2}_{T}}{4l^{2}_{L}}n^{2}),
\]
\begin{equation}
Z_{R}=\sum^{+\infty}_{n=1}\exp(-\beta E^{R}_{n})=\sum^{+\infty}_{n=1}\exp(-\frac{\pi\lambda^{2}_{T}}{4l^{2}_{R}}n^{2}),
\end{equation}
the density matrix $\rho_{\mathrm{eq}}$ is decomposed into 
\begin{equation}
\rho_{\mathrm{eq}}(l)=\frac{Z_{L}}{Z}\rho^{L}_{\mathrm{eq}}(l_{L})+\frac{Z_{R}}{Z}\rho^{R}_{\mathrm{eq}}(l_{R}).
\end{equation}

In the limit $l\gg\lambda_{T}$ , the density matrix $\rho_{\mathrm{eq}}$
takes the form 
\begin{equation}
\rho_{\mathrm{eq}}(l)|_{l\gg\lambda_{T}}=\frac{l_{L}}{l}\rho^{L}_{\mathrm{eq}}(l_{L})+\frac{l_{R}}{l}\rho^{R}_{\mathrm{eq}}(l_{R}).
\end{equation}

Equivalently, the method of projection operators can be used to decompose
the density matrix. Defining the projection operators 
\begin{equation}
P_{L}(l_{L})=\sum^{+\infty}_{n=1}\left|\psi^{L}_{n}\right\rangle \left\langle \psi^{L}_{n}\right|,P_{R}(l_{R})=\sum^{+\infty}_{n=1}\left|\psi^{R}_{n}\right\rangle \left\langle \psi^{R}_{n}\right|,
\end{equation}
we have $\rho_{\mathrm{eq}}(l)=P_{L}(l_{L})\rho_{\mathrm{eq}}(l)P_{L}(l_{L})+P_{R}(l_{R})\rho_{\mathrm{eq}}(l)P_{R}(l_{R})$,
where $(Z_{L}/Z)\rho^{L}_{\mathrm{eq}}(l_{L})=P_{L}(l_{L})\rho_{\mathrm{eq}}(l)P_{L}(l_{L}),(Z_{R}/Z)\rho^{R}_{\mathrm{eq}}(l_{R})=P_{R}(l_{R})\rho_{\mathrm{eq}}(l)P_{R}(l_{R})$.

Similar discussions can be extended to the case of multiple partitions.
When $k-1$ delta potentials are introduced into a one-dimensional
infinite potential well, projection operators can be defined for the
$k$ spatial regions $D_{1},D_{2},\cdots,D_{k}$ as $P_{i}(l_{i})=\sum^{+\infty}_{n=1}\left|\psi^{i}_{n}\right\rangle \left\langle \psi^{i}_{n}\right|,i=1,2,\cdots,k$,
where $\left|\psi^{i}_{n}\right\rangle $ are the eigenstates of the
one-dimensional infinite potential well corresponding to region $D_{i}$.
The density matrix can be decomposed in a similar manner as 
\begin{equation}
\rho_{\mathrm{eq}}(l)=\sum^{k}_{i=1}P_{i}(l_{i})\rho_{\mathrm{eq}}(l)P_{i}(l_{i}).
\end{equation}

\bibliographystyle{apsrev4-1}
\bibliography{Nonorthogonal}

\end{document}